\documentclass[english,aps,prl,twocolumn,superscriptaddress,showpacs]{revtex4}
\usepackage{graphicx}
\usepackage{float}
\usepackage{amssymb}
\usepackage{babel}

\begin{document}

\title{Global scale-invariant dissipation in collisionless plasma turbulence}

\author{K. H. Kiyani}

\email{k.kiyani@warwick.ac.uk}

\affiliation{Centre for Fusion, Space and Astrophysics; University of Warwick,
Coventry, CV4 7AL, United Kingdom}

\author{S. C. Chapman}

\affiliation{Centre for Fusion, Space and Astrophysics; University of Warwick,
Coventry, CV4 7AL, United Kingdom}

\author{Yu. V. Khotyaintsev}

\affiliation{Swedish Institute of Space Physics, Uppsala, Sweden}

\author{M. W. Dunlop}

\affiliation{Rutherford Appleton Laboratory, Didcot, United Kingdom}

\begin{abstract}
A higher-order multiscale analysis of the dissipation range of collisionless
plasma turbulence is presented using\emph{ in-situ} high-frequency
magnetic field measurements from the Cluster spacecraft in a stationary
interval of fast ambient solar wind. The observations, spanning five
decades in temporal scales, show a crossover from multifractal intermittent
turbulence in the inertial range to non-Gaussian monoscaling in the
dissipation range. This presents a strong observational constraint
on theories of dissipation mechanisms in turbulent collisionless plasmas.
\end{abstract}


\maketitle
The solar wind provides an ideal laboratory for the study of plasma
turbulence \citep{bruno2005}. \emph{In-situ} spacecraft observations
suggest well-developed turbulence at 1 AU with a magnetic Reynolds
number $\sim\mathcal{O}\left(10^{5}\right)$ \citep{Matthaeus2005,sorriso-valvo2007}.
These show an inertial range of Alfv\'enic turbulence on magnetohydrodynamic
(MHD) scales which is an anisotropic and possibly compressible energy
cascade \citep{Hnat2005b,Chapman2007,Horbury2008} with intermittent
magnetic field fluctuations described by statistical multifractals
and a power spectral density (PSD) with a scaling exponent close to
$-5/3$ \citep{bruno2005}. An outstanding problem is how, in the
absence of collisional viscosity in the solar wind, this inertial
range of MHD turbulence terminates at smaller scales where one anticipates
a cross-over to dissipative and/or dispersive processes via wave-particle
resonances. Understanding the nature of the dissipation processes
may also inform open questions such as how the solar wind and solar
coronal plasmas are heated \citep{Leamon2000,Kasper2008,araneda2009}. 

It has long been known \citep{Coleman1968,Behannon1978} that in collisionless
plasmas there is a transition in the PSD at high wavenumber $k$ from
MHD to kinetic physics at approximately the ion gyroscale. High resolution
\emph{in-situ} magnetic field observations reveal that at these scales
the turbulent solar wind shows a transition from a $\sim-5/3$ power
law in the inertial range to a steeper power-law at higher $k$ with
spectral exponents in the range $(-4,-2)$ \citep{Leamon1998b,smith2006}.
However, the relevant physical mechanism is much debated; having implications
for phenomena as diverse as magnetic reconnection \citep{Sundkvist2007,Eastwood2009},
neutron stars and accretion disks \citep{ChoLazarian2004}. Theories
which have been proposed range from nonlinear turbulent-like cascade
processes \citep{Galtier2007,Alexandrova2008c,Sahraoui2009} to weak
turbulence theories with wave dispersion and resonant plasma interactions
\citep{Leamon1998}. As well as studies of \emph{in-situ} spacecraft
measurements in the solar wind \citep{Bale2005}, foreshock \citep{Narita2006}
and magnetosheath \citep{Sahraoui2006,Alexandrova2008} regions, these
theories are explored using simulations ranging from Hall-MHD \citep{Shaikh2009},
electron-MHD \citep{Biskamp1996,ChoLazarian2004}, gyrokinetics \citep{Howes2008b},
particle-in-cell simulations of whistler turbulence \citep{Saito2008}
and Vlasov-hybrid simulations \citep{Valentini2008}. 
\begin{figure}
\begin{centering}
\includegraphics[clip,width=1\columnwidth,keepaspectratio]{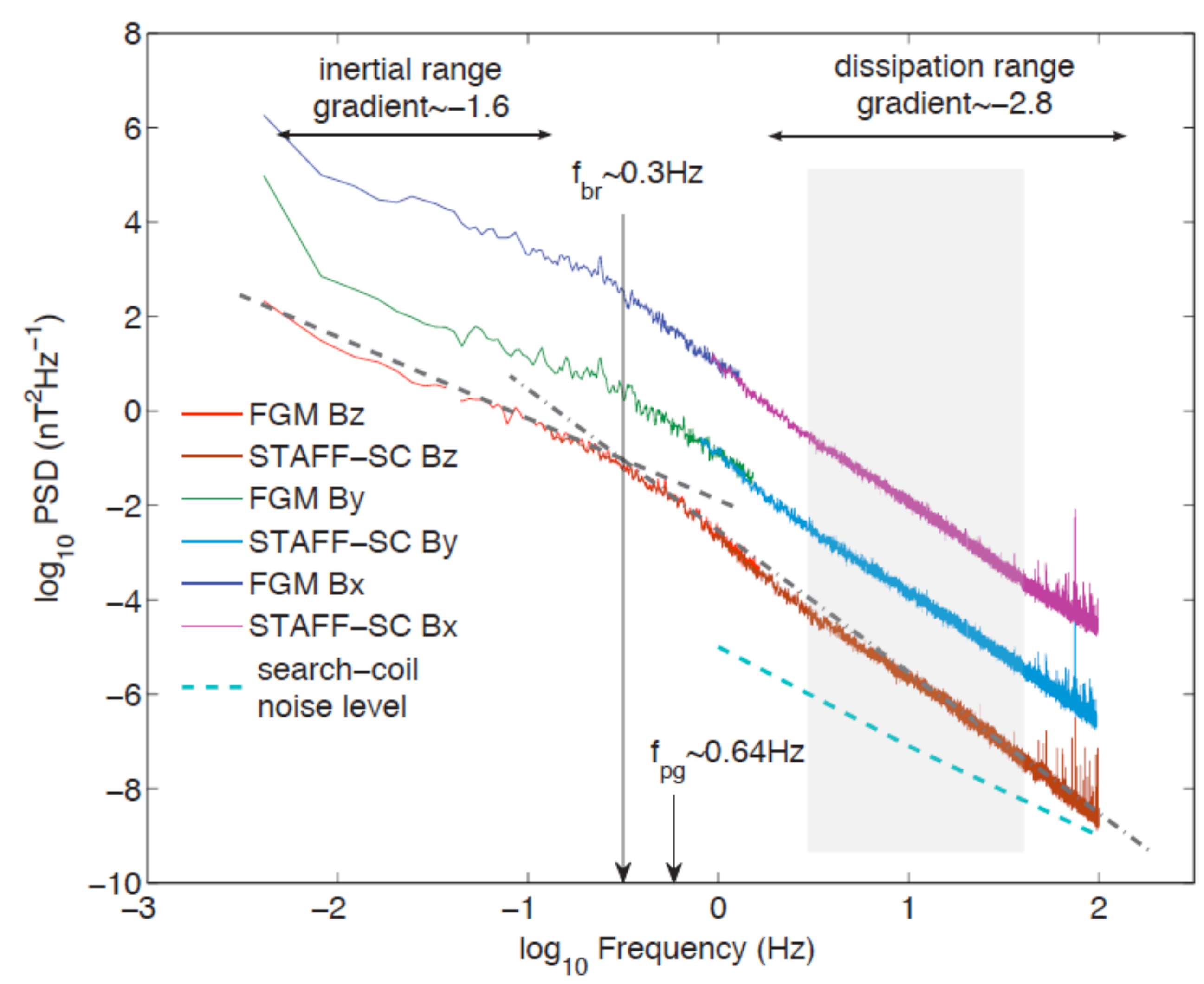}
\par\end{centering}

\caption{\label{fig:PSD}PSD plots of the components of the magnetic field
from both FGM (at frequencies lower than 1 Hz) and STAFF-SC (at frequencies
above 1Hz) instruments. The PSD values for $B_{x}$ and $B_{y}$ have
been shifted up for clarity. The 95\% confidence intervals for all
these plots are at $\pm0.03$ $log_{10}nT^{2}Hz^{-1}$. The shaded
region denotes the frequency range of STAFF-SC to be studied; the
upper limit corresponds to the search-coil signal-to-noise ratio (SNR)
of 10dB; the lower limit to the SNR of 20dB \citep{Pedersen1997}.
At lower frequencies the STAFF-SC response is attenuated by the instrument
callibration filter.}

\end{figure}

Both neutral fluid and MHD turbulence share a `classic' statistical
signature -- namely an intermittent multifractal scaling seen in the
higher-order statistics. In this letter we test the statistical properties
of the dissipation range and find in contrast monoscaling behaviour
i.e. a global scale-invariance. This provides a strong discriminator
for the physics and phenomenology of the dissipation range in collisionless
plasmas.

We present a detailed analysis of an interval of quiet, stationary
solar wind observed \emph{in-situ} by the Cluster spacecraft quartet
during an hour interval 00:10 -- 01:10 UT on January 30, 2007. By
combined analysis using high-frequency measurements of magnetic field
fluctuations from the search-coil (STAFF-SC) \citep{CWehrlin1997}
and flux-gate magnetometers (FGM) \citep{Balogh1997}, we simultaneously
probe both the inertial range and dissipation range up to frequencies
of 80 Hz. This will allow us to establish the scaling properties in
the dissipation range over about two decades in frequency. Figure
\ref{fig:PSD} shows the overlaid PSD from FGM and STAFF-SC for our
interval for the three magnetic field components $B_{x}$, $B_{y}$
and $B_{z}$ in geocentric ecliptic (GSE) coordinates -- both the
inertial and dissipation ranges can be clearly seen in all three components.
We indicate on the plot the frequency corresponding to the Doppler-shifted
proton gyroradius $f_{pg}$ \citep{Sahraoui2009} which is close to
the cross-over frequency break $f_{br}$ between the inertial and
dissipation range. 
\begin{figure}
\begin{centering}
\includegraphics[clip,width=1\columnwidth,keepaspectratio]{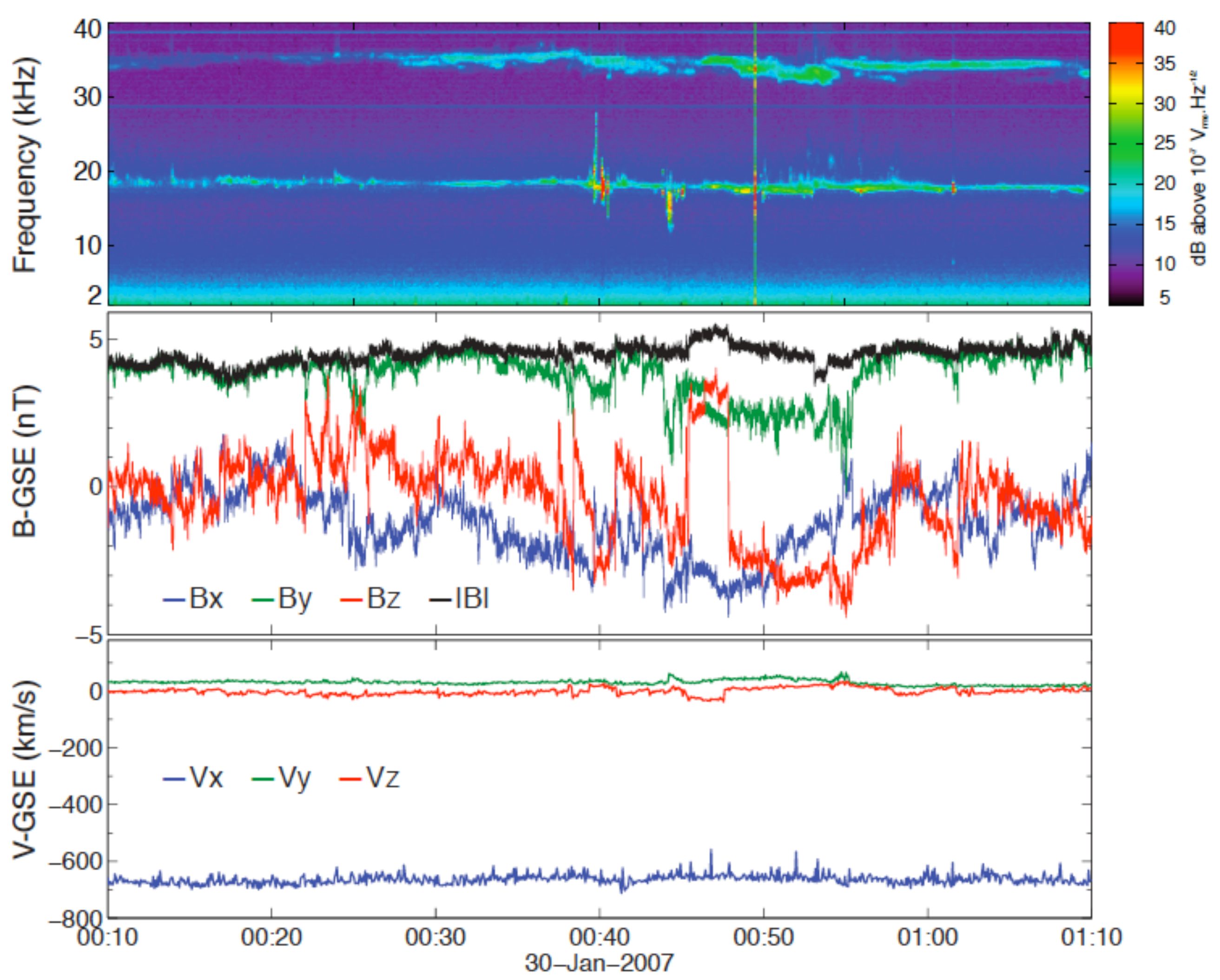}
\par\end{centering}

\caption{\label{fig:summary}Summary plot illustrating the quiet nature of
the solar wind interval under study. Top panel: E-field spectrogram
from WHISPER on Cluster spacecraft 4 showing steady plasma emissions;
center panel: B-field components; bottom panel: ion velocities from
HIA on Cluster spacecraft 1. }

\end{figure}
A summary of the solar wind interval used in this study is shown in
fig. \ref{fig:summary}. We plot the electric field spectrogram in
a frequency range 2 -- 40 kHz from WHISPER \citep{Decreau1997}, the
magnetic field timeseries from FGM and the ion velocity from CIS/HIA
\citep{Reme1997} during the interval of interest. The constant ion
velocity and plasma density (evidenced by the constant electron plasma
frequency on the E-field spectrogram) indicate a stationary pristine
interval of ambient solar wind disconnected from the Earth's bowshock,
away from ion and electron foreshock regions and imbedded in a steady
fast solar wind stream $\gtrsim650\, km\, s^{-1}$ \emph{i.e.} quintessential
plasma turbulence away from any external \emph{a priori} physical
processes and dynamics. Due to stationarity of plasma parameters and
the B-field magnitude it is sufficient to quote single values for
the other relevant parameters: ion temperature $T_{i}\simeq1.2\, MK$,
Alfv\'en speed $V_{A}\simeq50\, km\, s^{-1}$, ion plasma beta $\beta\simeq2$,
and plasma density $n_{e}\simeq3.8\, cm^{-3}$ (from the plasma frequency
measured by WHISPER).

Importantly for this study, we have chosen an interval where both
STAFF-SC and FGM were operating in \emph{burst-mode} so that one can
access the largest range of scales available and probe further into
smaller scales where the dissipation range can be studied. For this
interval STAFF-SC provides AC waveform data in a frequency band between
$\sim0.1$ Hz and $180$ Hz (low-pass filter); our study is restricted
to frequencies lower than 80 Hz to maintain good SNR. FGM provides
DC waveform data with largest frequency at $\sim33$ Hz (Nyquist cutoff).
This ensures a large sample ($\sim1.6\times10^{6}$ for STAFF-SC and$\sim2.5\times10^{5}$
for FGM) and thus well-resolved statistics. The Welch PSD method used
in fig. \ref{fig:PSD} employs 50\% overlapped windows each containing
a sample of $2^{14}$ points for FGM and $2^{16}$ points for STAFF-SC;
this provides a very large range of frequencies to study as well as
reducing the error due to noise on the PSD measurements. The large
sample size coupled with the intervals stationarity ensures that we
have a good control of errors which can arise due to finite sample
size \citep{Kiyani2009a}. All the results presented in this paper
are from Cluster spacecraft 4; the $B_{z}$ component provides the
largest overlap in frequencies between both FGM and STAFF-SC due to
it being out of the spin plane of the spacecraft and thus is cleanest
with respect to spin tone contamination. As we will be studying single
point Eulerian measurements in very fast solar wind, streaming past
the spacecraft providing a single time-series, we will assume the
validity of Taylor's frozen-in-turbulence hypothesis \citep{Sahraoui2006}
which uses time as a proxy for space -- although our PSD and other
statistics will always be presented in the frequency and time domains.

The PSD provides one statistic to probe the scale dependent behaviour
of the turbulent fluctuations and is equivalent to studying the autocorrelation
-- a second order statistic. To test for multifractal scaling of the
fluctuations we now turn to higher-order statistics. We focus on the
statistics of magnetic field increments defined as $\delta B_{i}(t,\tau)=B_{i}(t+\tau)-B_{i}(t)$
for each vector component $i$ and time lag or scale $\tau$; in particular
the focus is on the absolute moments of these increments, also known
as structure functions
\begin{equation}
S_{i}^{m}(\tau)=\frac{1}{N}\sum_{j=1}^{N}\left|\delta B_{i}(t_{j},\tau)\right|^{m}\ ,\label{eq:1}
\end{equation}
where $m$ is the moment order. We have verified the statistical stationarity
of the interval being studied and thus can form an ensemble average
by taking a time average (assuming ergodicity) along the signal of
sample size $N$. Importantly, the higher-order structure functions
progressively capture the more intermittent, larger fluctuations.
As we are studying the magnetic field increments, these large fluctuations
represent the spatial gradients which are responsible for dissipating
energy from the magnetic fields. We will focus on the scaling behaviour
of the structure functions with scale $\tau$ such that
\begin{equation}
S_{i}^{m}(\tau)\propto\tau^{\zeta(m)}\ ,\label{eq:2}
\end{equation}
where linear dependence of the scaling exponent $\zeta(m)=Hm$ implies
monoscaling with a single exponent $H$. In theories of turbulence
non-linear $\zeta(m)$ behaviour is associated with the intensity
of energy dissipation being distributed on a spatial multifractal
\citep{bruno2005}. 

\begin{figure}
\begin{centering}
\includegraphics[clip,width=1\columnwidth,keepaspectratio]{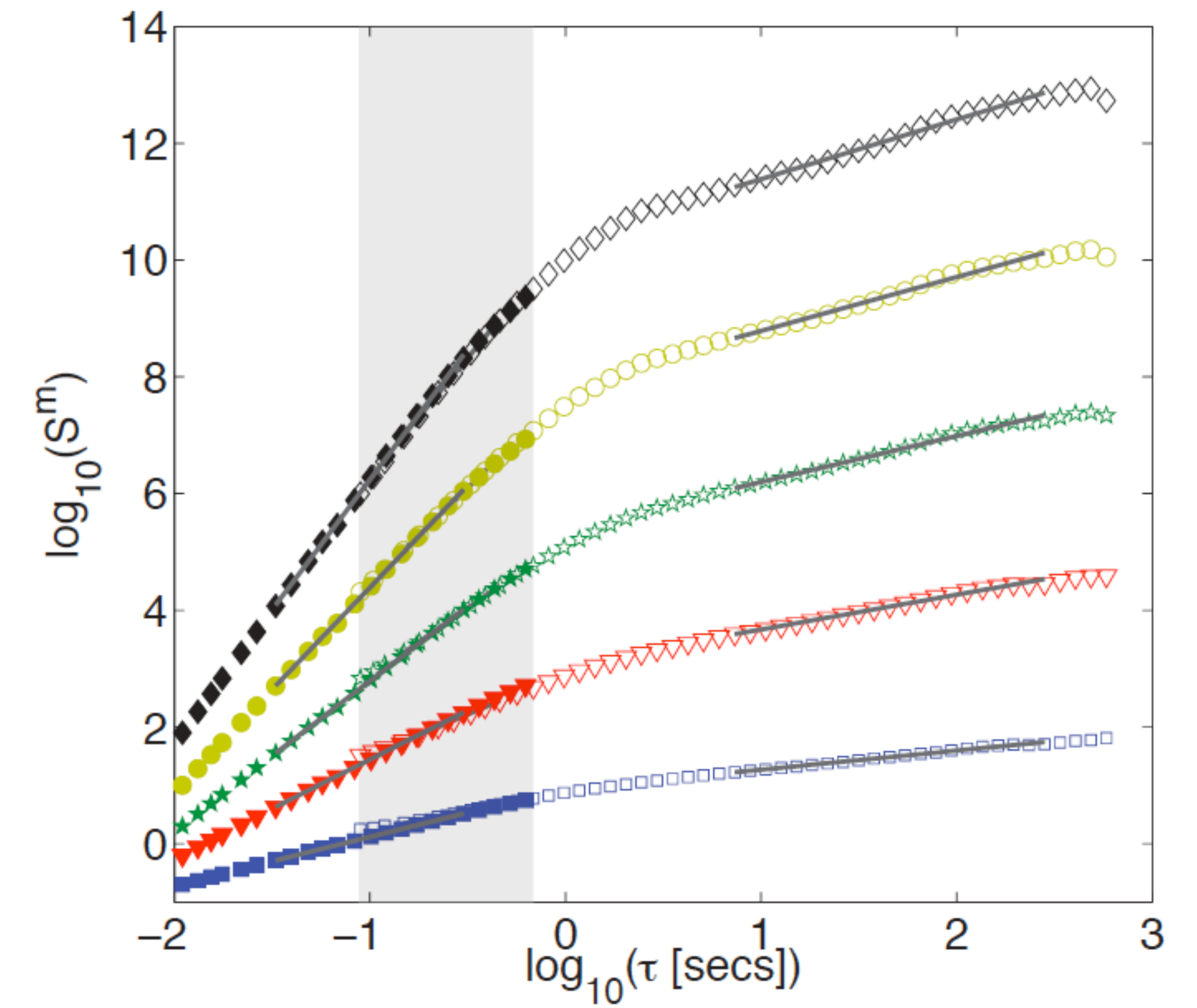}
\par\end{centering}

\caption{\label{fig:SF}Structure functions of orders: 1-$\square$, 2-$\triangledown$,
3-$\star$, 4-$\circ$ and 5-$\lozenge$. Open shapes correspond to
FGM measurements and filled shapes refer to STAFF-SC. The curves have
been shifted along the vertical axis to allow a comparison of the
gradients. The shaded area indicates the scales where both FGM and
STAFF overlap. Linear fits for the inertial and dissipation ranges
are also shown.}

\end{figure}
\begin{figure}
\begin{centering}
\includegraphics[clip,width=1\columnwidth,keepaspectratio]{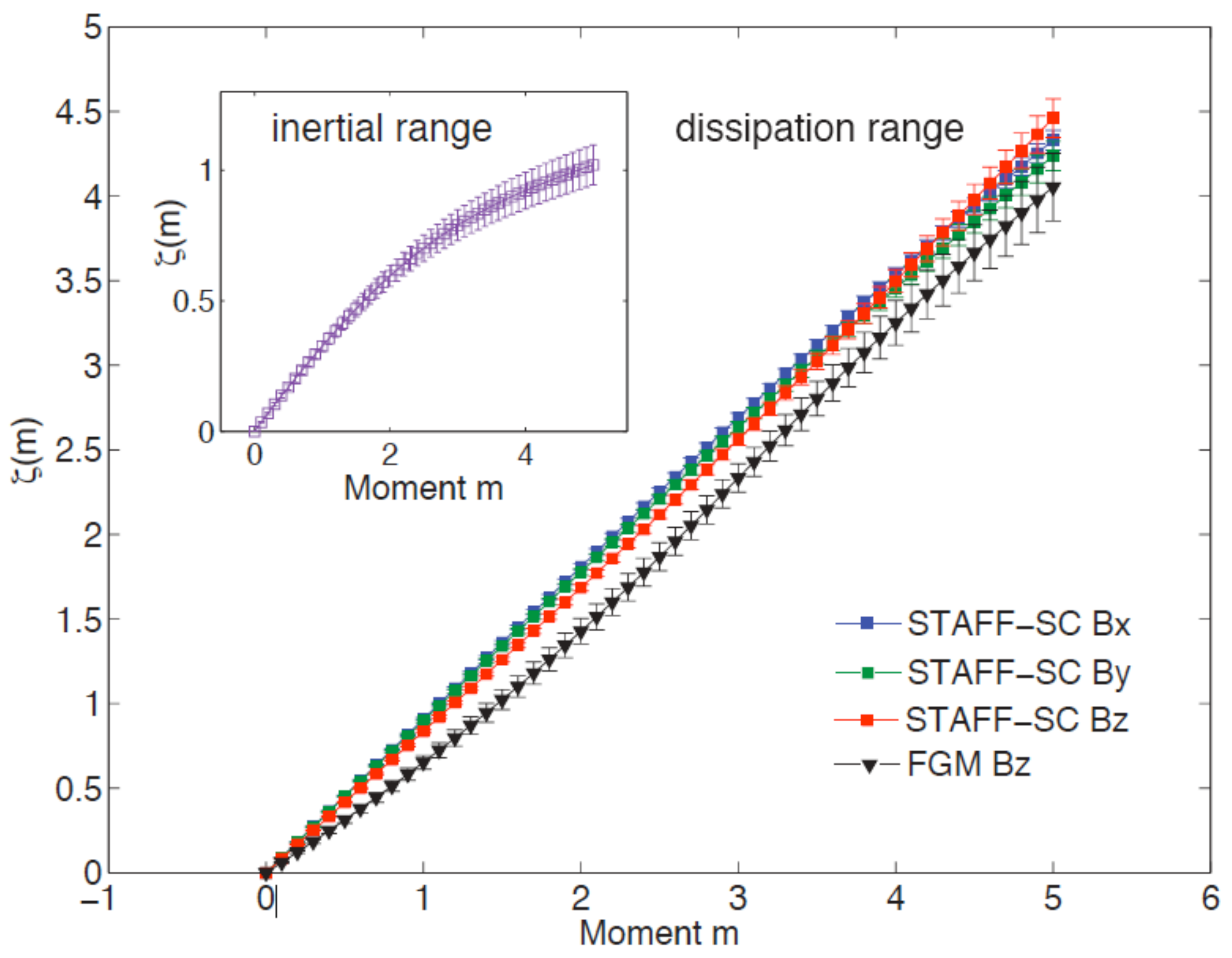}
\par\end{centering}

\caption{\label{fig:zeta}Main plot: Scaling exponents $\zeta$ with order
$m$; a linear relationship on this plot indicates monoscaling behaviour.
$\zeta(m)$ obtained from both FGM and STAFF-SC are shown for $B_{z}$;
these show close correspondence. STAFF-SC $B_{x}$, $B_{y}$ components
are also shown and indicate isotropic scaling. Inset: $\zeta(m)$
Vs. $m$ for the inertial range using FGM $B_{z}$; this is concave,
consistent with the multifractal nature of the inertial range.}

\end{figure}

The structure functions for the data interval studied here are shown
in fig. \ref{fig:SF}. On this $log-log$ plot the gradients as shown
give estimates of the scaling exponents $\zeta(m)$. The inertial
and dissipation ranges of scaling are well-defined with a sharp transition
at the break point at $\simeq3$ seconds in agreement with the PSD
in fig. \ref{fig:PSD}; the dissipation range extends over nearly
two orders of magnitude. Importantly, there is excellent agreement
between STAFF-SC and FGM in the dissipation range where they overlap
for almost a decade, indicated by the shaded region on the plot. We
plot $\zeta(m)$ vs. $m$ for the dissipation range in the main panel
of fig. \ref{fig:zeta}, and for the inertial range in the inset.
Surprisingly, the dissipation range is monoscaling i.e. globally scale-invariant;
in contrast to the inertial range which is multifractal, characteristic
of fully developed turbulence with $\zeta(2)\sim2/3$. The single
scaling parameter for the dissipation range for $B_{z}$ is $H=0.89\pm0.02$
for STAFF-SC and $H=0.84\pm0.05$ for FGM. To test the robustness
of this result we have repeated this analysis for another ambient
fast solar wind interval (12:10 -- 14:00 UT January 20, 2007) and
obtained the same global scale-invariance. In both cases we find that
all three field components are monoscaling. We can see that for the
particular solar wind interval shown in fig. \ref{fig:zeta} the exponents
$H$ for $B_{x}$ and $B_{y}$ are close to that of $B_{z}$, suggesting
that the small scale features of this turbulent interval of the solar
wind are also isotropic. For STAFF-SC data from the second interval,
however, we find $H=0.9\pm0.02$ for $B_{x}$ and $B_{y}$ and $H=0.8\pm0.05$
for $B_{z}$, suggesting an anisotropy that may depend on local plasma
parameters. 

Monoscaling of the structure functions implies that the probability
density function (PDF) of the increments $P\left(\delta B_{i},\tau\right)$
at a particular scale $\tau$ should collapse onto a unique scaling
function $\mathcal{P}_{s}$ via the following rescaling operation
\citep{Kiyani2006} 
\begin{equation}
\mathcal{P}_{s}(\delta B_{i}\tau^{-H})=\tau^{H}P\left(\delta B_{i},\tau\right)\ .\label{eq:3}
\end{equation}
\begin{figure}
\begin{centering}
\includegraphics[clip,width=1\columnwidth,keepaspectratio]{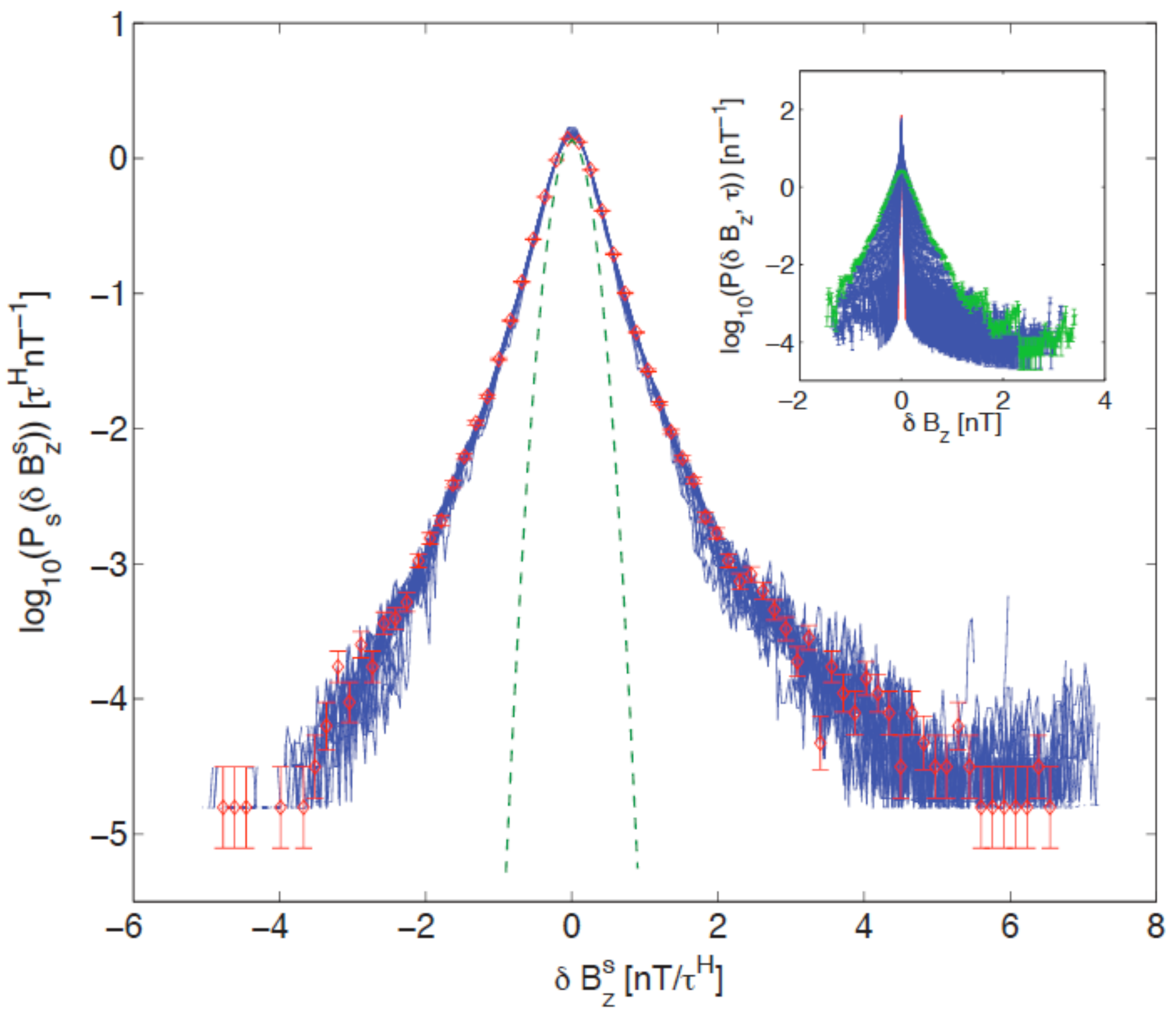}
\par\end{centering}

\caption{\label{fig:PDF}Main plot: PDFs rescaled using eq. (\ref{eq:3}) $\left(\delta B_{z}^{s}=\delta B_{z}\tau^{-H}\right)$.
Inset: PDFs at different scales $\tau$ before rescaling; red and
green curves show the smallest and largest values of $\tau$ respectively.
A Gaussian fit to the data illustrates the heavy-tailed non-Gaussian
nature of the rescaled PDF.}
\end{figure}
This collapse of the data to a single scaling function is tested
in fig. \ref{fig:PDF} for $B_{z}$, where we have used the same values
of $\tau$ and the $H$ value obtained above. We can see that there
is an excellent collapse onto a single curve. A fitted Gaussian illustrates
the highly non-Gaussian nature of the tails of this PDF. 

In conclusion, our results suggest that global scale invariance in
small-scale magnetic fluctuations is a robust feature of the dissipation
range of collisionless plasma turbulence in the fast ambient solar
wind. This is a surprising result as it is distinct from the multifractal
scaling that is characteristic of both neutral fluid and MHD turbulent
cascades in the inertial range. Successful theoretical understanding
of the dissipation range should include this property. Our result
provides a strong discriminator of the relevant physics and phenomenology;
for example the monoscaling that we find is reminiscent of that found
at higher orders in electron-MHD simulations \citep{Boffetta1999}.
To determine whether this phenomenology is in fact universal, future
studies should aim to reproduce and/or break this result in more dynamic
environments such as at planetary shocks \citep{Narita2006}, magnetosheath
\citep{Alexandrova2008} and at sites of magnetic reconnection \citep{Sundkvist2007,Eastwood2009};
although the main difficulty here will be to identify sufficiently
long stationary intervals. 

\begin{acknowledgments}
We thank E. Yordonova, C. Foullon, B. Hnat, ISSI team 132 and in particular
T. Dudok De Wit for their help and suggestions; P. Canu and O. Santolik
for STAFF-SC calibration queries; N. Cornilleau-Wehrlin, A. Balogh,
H. R\'eme and their teams for use of STAFF-SC, FGM and CIS/HIA data
respectively; and P. D\'ecr\'eau and S. Grimald for the WHISPER
spectrogram. This work was supported by the UK STFC.
\end{acknowledgments}

\bibliographystyle{apsrev}
%\bibliography{PRLKKSmScPlTur}

\end{document}